\documentclass[aps,preprint,prd,showpacs,nofootinbib]{revtex4}

\usepackage{amsmath,amssymb}
\usepackage{graphicx,subfigure}
\usepackage{color}
\usepackage[colorlinks,linkcolor=magenta,anchorcolor=cyan,citecolor=blue,plainpages=false]{hyperref}

\def\lf{\left}
\def\rt{\right}

\def\be{\begin{equation}}
\def\ee{\end{equation}}
\def\ba{\begin{eqnarray}}
\def\ea{\end{eqnarray}}

\begin{document}

    \title{Testing dark energy after pre-recombination early dark energy}

    \author{Hao Wang$^{1,2} $\footnote{\href{wanghao187@mails.ucas.ac.cn}{wanghao187@mails.ucas.ac.cn}}}

    \author{Yun-Song Piao$^{1,2,3,4} $ \footnote{\href{yspiao@ucas.ac.cn}{yspiao@ucas.ac.cn}}}

\affiliation{$^1$ School of Fundamental Physics and Mathematical
    Sciences, Hangzhou Institute for Advanced Study, UCAS, Hangzhou
    310024, China}

\affiliation{$^2$ School of Physics Sciences, University of
Chinese Academy of Sciences, Beijing 100049, China}

\affiliation{$^3$ International Center for Theoretical Physics
    Asia-Pacific, Beijing/Hangzhou, China}

\affiliation{$^4$ Institute of Theoretical Physics, Chinese
Academy of Sciences, P.O. Box 2735, Beijing 100190, China}

    \begin{abstract}

In the studies on pre-recombination early dark energy (EDE), the
evolution of Universe after recombination is usually regarded as
$\Lambda$CDM-like, which corresponds that the equation of state of
dark energy responsible for current accelerated expansion is
$w=-1$. However, in realistic models, $w$ might be evolving. We
consider the parametrizations of $w$ with respect to the redshift
$z$ in Axion-like EDE and AdS-EDE models, respectively. We
performed the Monte Carlo Markov chain analysis with recent
cosmological data, and found that the bestfit $w(z)$ is compatible
with $w_0=-1,w_a=0$ (the cosmological constant) and the evolution
of $w$ is only marginally favored, which so has little effect on
lifting the bestfit value of ${H_0}$.

    \end{abstract}

    \thispagestyle{empty}
    \maketitle
    \newpage

\section{Introduction}
Recently, local measurements of Hubble constant $H_0$ have showed
$H_0\sim 73$km/s/Mpc
\cite{Riess:2019cxk,Chen:2019ejq,Wong:2019kwg,Freedman:2019jwv,Huang:2019yhh},
which has $\gtrsim 4\sigma$ discrepancy to $H_0\sim 67$km/s/Mpc
suggested from the fit of cosmic microwave background (CMB) data
\cite{Planck:2018vyg} based on the $\Lambda$CDM model, which is
so-called Hubble tension \cite{Riess:2019qba,Verde:2019ivm}, see
also the most recent \cite{Riess:2021jrx}. Currently, it is
difficult to explain this conflict by systematic errors.

It has been widely thought that the Hubble tension might be a hint
of new physics beyond $\Lambda$CDM,
e.g.\cite{Knox:2019rjx,Cai:2021weh,Lyu:2020lwm,Haridasu:2020pms,Vagnozzi:2021gjh}.
In early dark energy (EDE) scenario \cite{Poulin:2018cxd}, see
also
\cite{Agrawal:2019lmo,Alexander:2019rsc,Lin:2019qug,Niedermann:2019olb,Ye:2020btb,Ye:2020oix,Smith:2019ihp,Sakstein:2019fmf,Lin:2020jcb,Braglia:2020bym,Karwal:2021vpk,Nojiri:2021dze,Sabla:2021nfy,Moss:2021obd,McDonough:2021pdg,Gomez-Valent:2021cbe},
since the scalar field energy has a non-negligible fraction before
recombination, the sound horizon \be
r^*_s=\int_{z_*}^\infty{c_s\over H(z)} dz \ee
 is suppressed, where $z_*$ is the redshift
at recombination, which naturally brings a higher $H_0$ (noting
that CMB and BAO have fixed the angular scales
\be
\theta^*_s={r_s^*\over D_A^*}\sim r_s^* H_0,
\ee
 where $D_A^*$ is the angular
diameter to last scattering surface), without spoiling fit to CMB
and baryon acoustic oscillations (BAO) data, see also
\cite{LaPosta:2021pgm,Chudaykin:2020acu,Chudaykin:2020igl,Jiang:2021bab}
for combined Planck+SPT dataset and
\cite{Hill:2021yec,Poulin:2021bjr} for Planck+ACT dataset. In
particular, an Anti-de Sitter (AdS) phase around recombination can
allow $H_0\simeq 73$km/s/Mpc, so the corresponding AdS-EDE model
\cite{Ye:2020btb,Ye:2020oix} can be $1\sigma$ consistent with
local $H_0$ measurements. In Ref.\cite{Ye:2021nej}, it has been
found that the pre-recombination solutions of the Hubble tension
implies a scale-invariant Harrison-Zeldovich spectrum of
primordial scalar perturbation, i.e. $n_s= 1$ for $H_0\sim
73$km/s/Mpc.

The beyond-$\Lambda$CDM modifications after
recombination have also been proposed
e.g.\cite{DiValentino:2017zyq,Vagnozzi:2019ezj,Visinelli:2019qqu,DiValentino:2019ffd,Akarsu:2019hmw,Akarsu:2021fol,Yang:2021flj,Yang:2020ope,Liu:2021mkv,Alestas:2021luu,Perivolaropoulos:2021bds,Bag:2021cqm},
see also \cite{DiValentino:2021izs,Perivolaropoulos:2021jda} for
recent reviews. The current accelerated expansion of our Universe
suggests the existence of dark energy at present, with the
equation of state (EOS) $w={p/\rho}\simeq -1$. In the studies on
pre-recombination EDE, the evolution of Universe after
recombination is usually regarded as $\Lambda$CDM-like, which
corresponds to $w=-1$. However, in realistic models, $w$ might be
evolving, see e.g.\cite{DiValentino:2021izs} for a review, so
having a consistency check for the $\Lambda$CDM model after
recombination is significant.

In this paper, we will consider different parametrization of $w$
in Axion-like EDE and AdS-EDE models, respectively, and perform
the Markov Chain Monte Carlo (MCMC) analysis with PlanckCMB, BAO,
Pantheon and $H_0$ dataset. This paper is organised as follows. In
sect-II, we outline the parametrizations of $w$ of current dark
energy. In sect-III, we perform the MCMC analysis and present our
results. We conclude in sect-IV.

\section{Parametrizations of dark energy}

In $\Lambda$CDM model, the dark energy activates as a cosmological
constant with $w=-1$. However, in scalar field models of dark
energy,
e.g.\cite{Ratra:1987rm,Zlatev:1998tr,Caldwell:1999ew,Feng:2004ad,Guo:2004fq,Wei:2005nw,Wei:2005fq},
$w$ may be evolving or oscillating. Here, it is convenient to work
with the parametrizations of $w(z)$.

The simplest parametrization is $w(z)=w_0+{w_a}z$, which is linear
but diverge at high redshift, see also
\cite{Cooray:1999da,Astier:2000as,Weller:2001gf}. Thus the CPL
parametrization \cite{Chevallier:2000qy,Linder:2002et}
    \be
    w(z)=w_0+{w_a}\lf(\frac{z}{1+z}\rt),
    \label{CPL}
    \ee
and the oscillating parametrization
$w(z)=w_0+w_a\{1-\cos\lf[\ln(1+z)\rt]\}$ \cite{Feng:2004ff}, see
also \cite{Lazkoz:2010gz,Pan:2017zoh}, have been widely applied.
Both are well behaved at high $z$. The parametrizations degrading
into linear at low $z$ also include
logarithmic\cite{Efstathiou:1999tm}, etc
\cite{Barboza:2008rh,Ma:2011nc,Feng:2012gf,Pantazis:2016nky,Yang:2017alx}, see also \cite{Colgain:2021pmf} for relevant comments.

\section{MCMC results for EDE models}

In this section we will confront the Axion EDE and AdS-EDE models
(with CPL and oscillating parametrizations, respectively) with
recent cosmological data. We modified the Montepython-3.3
\cite{Audren:2012wb,Brinckmann:2018cvx} and CLASS
\cite{Lesgourgues:2011re,Blas:2011rf} codes to perform the MCMC
analysis.

Here, we will set the SH0ES result \cite{Riess:2019cxk} as the
Gaussian prior. The datasets consist of Planck2018 high-l and
low-l TTTEEE as well as Planck lensing
likelihoods\cite{Planck:2019nip}, the BOSS DR12
\cite{BOSS:2016wmc} with its full covariant matrix for BAO as well
as the 6dFGS \cite{Beutler:2011hx} and MGS of SDSS
\cite{Ross:2014qpa} for low-$z$ BAO, the Pantheon data
\cite{Pan-STARRS1:2017jku}. We consider chains to be converged when the Gelman-Rubin statistic\cite{Gelman:1992zz} satisfies $R-1<0.05$.

\subsection{Axion-like EDE model \protect \footnote{We follow the name in Ref.\cite{Poulin:2021bjr}.}}
In Axion-like EDE model \cite{Poulin:2018cxd}, an oscillating axion
field with the potential $V(\phi)\propto(1-cos[\phi/f])^n$,
naturally arising in the string theory, is responsible for EDE. At
the critical resdshift, EDE starts to oscillate and dilutes away
like a fluid with $w=(n-1)/(n+1)$. It is noted that $n=3$ is
better for a higher best-fit $H_0$
\cite{Poulin:2018cxd,Smith:2019ihp}.

\begin{figure}[htbp]
   {\includegraphics[width=0.5\textwidth]{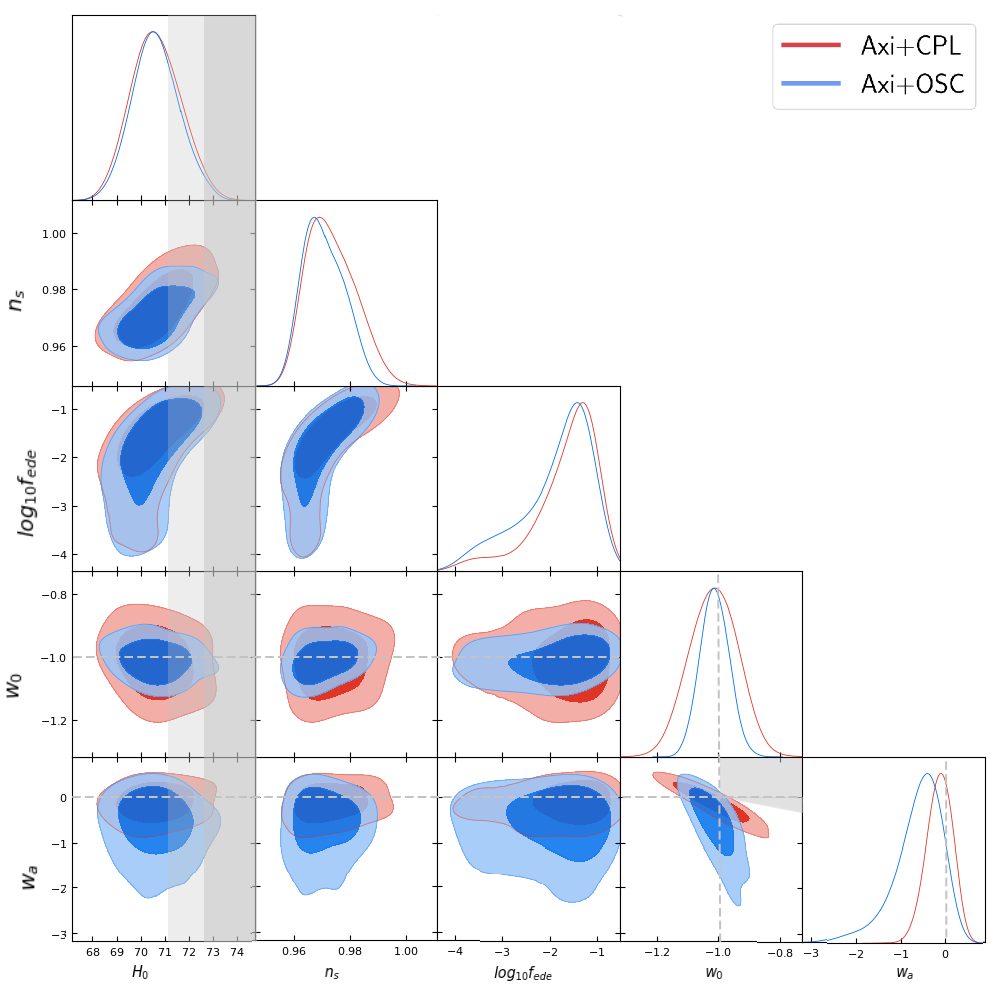}}\label{Axi}
\caption{\label{fig1}Marginalized $1\sigma$ and $2\sigma$ posterior
distributions of \{$H_0$, $n_s$,$f_{ede}$,$w_0$, $w_a$\} in Axion-like
EDE model. The red lines are the results for the CPL
parameterization and the blue ones are for the oscillating
parameterization. The shadow regions of $H_0$ represent 1$\sigma$ and 2$\sigma$
values of local measurement \cite{Riess:2019cxk}. The intersection
point of dash lines in $w_0$-$w_a$ plot corresponds to the
${\Lambda}$CDM model, with the shadow region corresponding to the
Quintessence dark energy in the whole history. }
    \end{figure}

We perform the MCMC analysis on the parameters set
$\{\omega_b,\omega_{cdm},H_0,\ln({10^{10}}A_s),n_s,\tau_{reio},{\log_{10}}{a_c},f_{ede},\phi_i,w_0,w_a\}$,
where $\phi_i$ is the initial value of EDE field, $a_c$ is when
the field starts rolling and $f_{ede}$ is the energy fraction of
EDE at $a_c$. We also set $n=3$
\cite{Poulin:2018cxd,Smith:2019ihp}. In Table-I, we present the
MCMC results for the CPL and oscillating parametrizations of
$w(z)$, respectively. In Fig.\ref{fig1}, we show the 1$\sigma$ and
2$\sigma$ marginalized posterior distributions of model
parameters.

\begin{table}[htbp]
    \begin{tabular}{c|c|cc}
        \hline
        Parameters&$\Lambda$CDM&Axi+CPL&Axi+OSC\\
        \hline
        100$\omega_b$&$2.247(2.224)^{+0.015}_{-0.014}$&$2.247(2.289)^{+0.0188}_{-0.0280}$&$2.256(2.277)^{+0.0256}_{-0.0267}$\\
        $\omega_{cdm}$&$0.1182(0.1183)^{+0.0008}_{-0.0013}$&$0.1243(0.1305)^{+0.00239}_{-0.00520}$&$0.1261(0.1307)^{0.00327}_{-0.00480}$\\
        $H_0$&$68.16(68.23)^{+0.56}_{-0.4}$&$70.06(72.05)^{+0.859}_{-1.346}$&$70.09(71.90)^{+1.182}_{-0.941}$\\
        ln($10^{10}$$A_s$)&$3.049(3.054)^{+0.013}_{-0.016}$&$3.044(3.050)^{+0.0172}_{-0.0148}$&$3.044(3.031)^{+0.0154}_{-0.0153}$\\
        $n_s$&$0.9688(0.9696)^{+0.0039}_{-0.0042}$&$0.9717(0.9853)^{+0.00069}_{-0.00104}$&$0.9749(0.9850)^{+0.00815}_{-0.01103}$\\
        $\tau_{reio}$&$0.0604(0.0636)^{+0.0066}_{-0.0075}$&$0.0525(0.0505)^{+0.00807}_{-0.00815}$&$0.0516(0.0448)^{+0.00721}_{-0.00706}$\\
        $f_{ede}$&-&$0.045(0.052)^{+0.0713}_{-0.0370}$&$0.047(0.054)^{+0.0794}_{-0.0229}$\\
        log$_{10}{a_c}$&-&$-3.652(-3.839)^{+0.2439}_{-0.2057}$&$-3.664(-3.864)^{+0.2386}_{-0.2112}$\\
        \hline
        100$\theta_s$&$1.0422(1.0421)^{+0.0005}_{-0.0004}$&$1.0416(1.0415)^{+0.000342}_{-0.000327}$&$1.0415(1.0415)^{+0.000338}_{-0.000379}$\\
        $\sigma_8$&$0.8078(0.81)^{+0.0054}_{-0.0066}$&$0.8472(0.8597)^{+0.0105}_{-0.0132}$&$0.8506(0.8538)^{+0.0115}_{-0.0119}$\\
        \hline
        $w_0$&-1&$-1.021(-0.970)^{+0.0755}_{-0.0883}$&$-1.008(-1.007)^{+0.0458}_{-0.0449}$\\
        $w_a$&0&$-0.130(-0.293)^{+0.347}_{-0.280}$&$-0.469(-0.444)^{+0.567}_{-0.322}$\\
        \hline
    \end{tabular}
\caption{Mean values (best-fits) of parameters in Axion-like EDE
model for CPL and oscillating parameterizations, respectively,
fitted to Planck2018+BAO+Pantheon+$H_0$ dataset.}
    \label{tab2}
\end{table}

In Fig.\ref{fig1}, we see that the evolving $w$ has little effect
on ${H_0}$ and $n_s$, compared with the model with $w=-1$ in
Ref.\cite{Poulin:2018cxd}. However, the EDE parameters are
constrained more tightly by the oscillating parametrization. The
result on $w(z)$ is compatible with the cosmological constant
($w_0=-1,w_a=0$), and only marginally favors the evolution of $w$.
Here and in AdS-EDE model (see Table-II), the amplitude $\sigma_8$
of matter density fluctuation at low redshift is larger than local
measurements \cite{Hildebrandt:2018yau,Heymans:2020gsg}, which,
however, may be pulled lower by new physics beyond cold dark
matter
\cite{Allali:2021azp,Ye:2021iwa,Clark:2021hlo}\footnote{Here, we
will not involve it.}.

\begin{figure}[htbp]
    \subfigure{\includegraphics[width=0.45\textwidth]{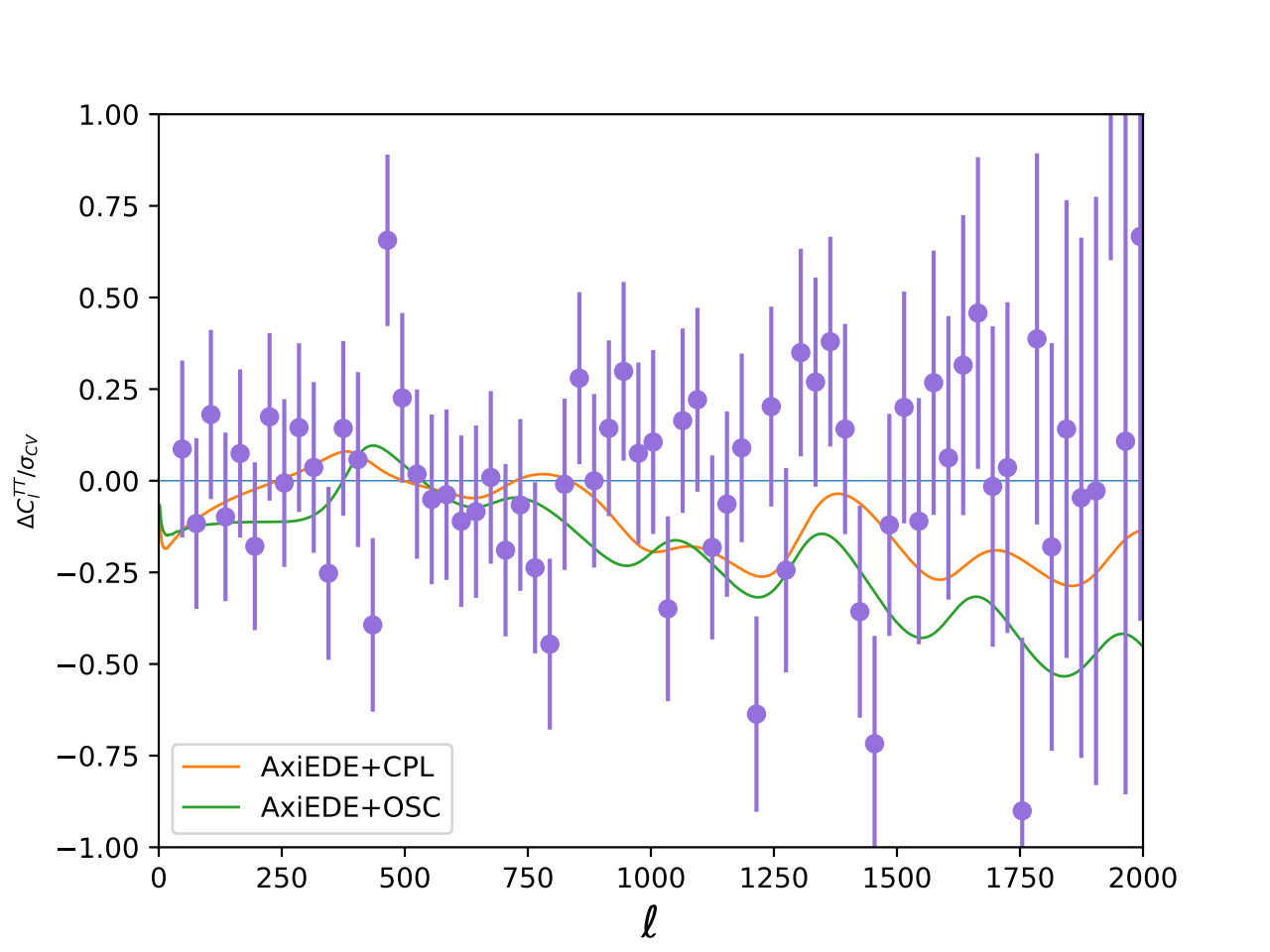}}\label{AxiTT}
    \subfigure{\includegraphics[width=0.45\textwidth]{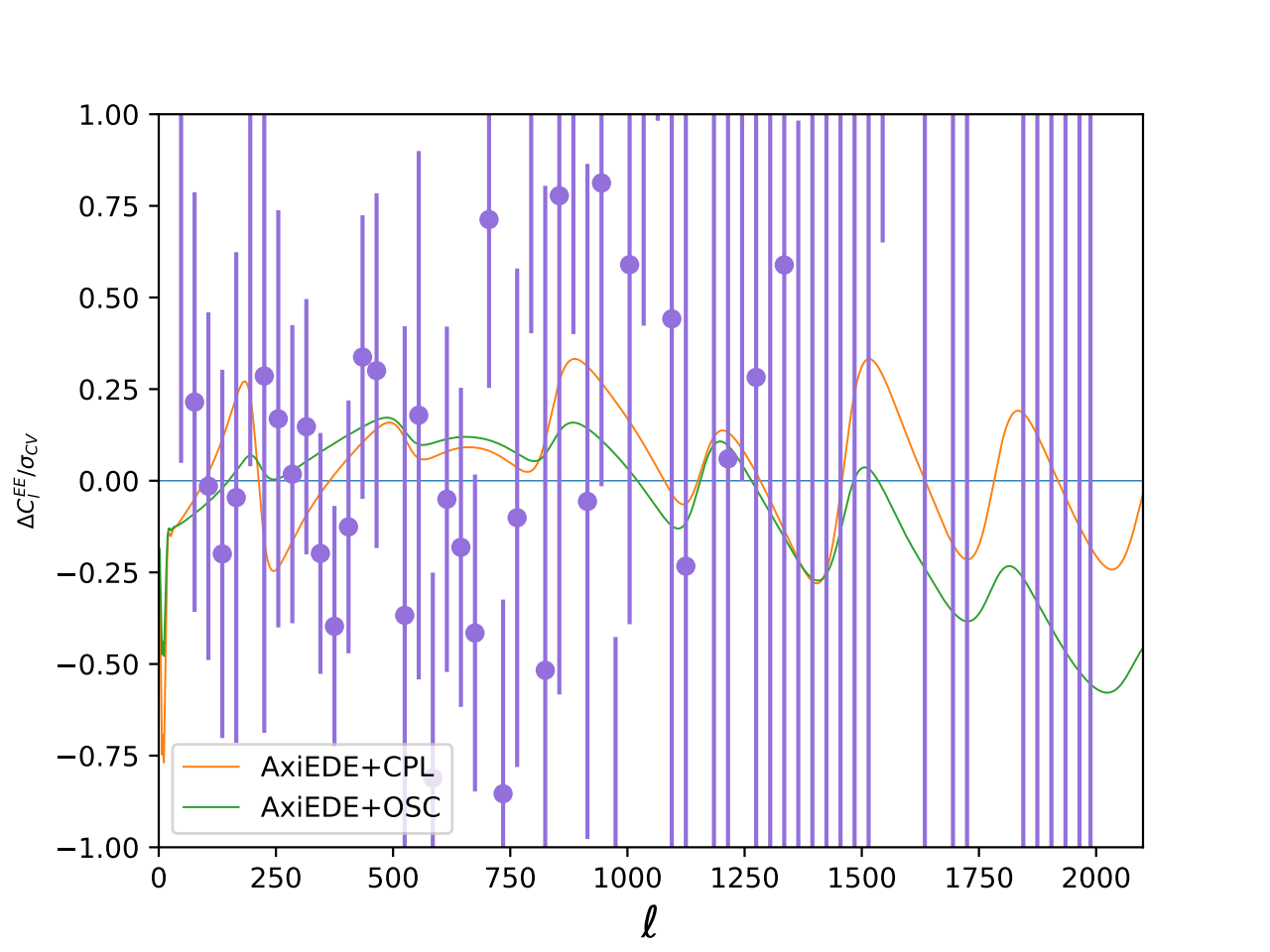}}\label{AxiEE}
    \subfigure{\includegraphics[width=0.45\textwidth]{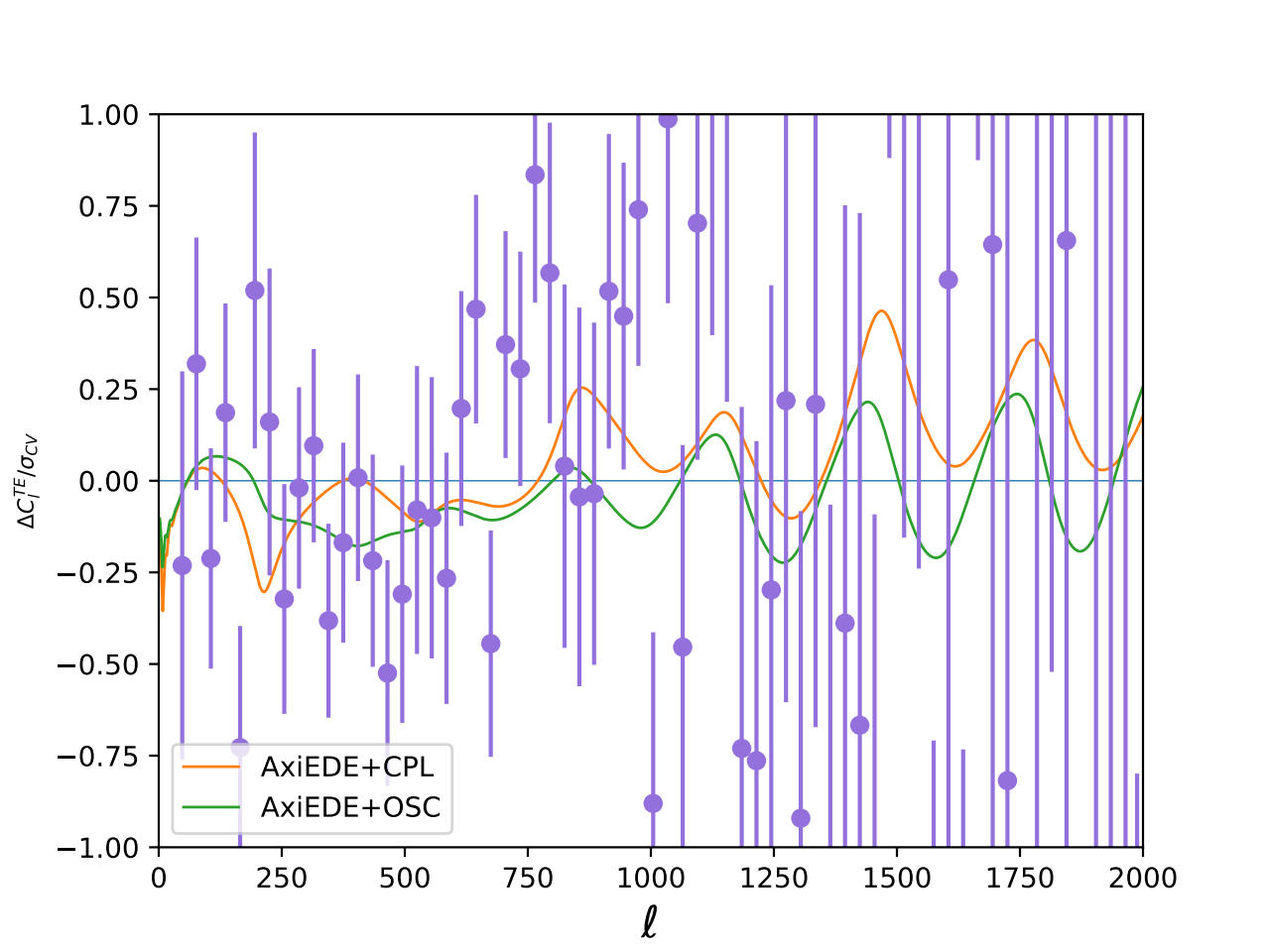}}\label{AxiTE}
\caption{\label{fig2}${\Delta}C_l/\sigma_{CV}$ for both parametrizations in
Axion-like EDE model fitted to Planck2018+BAO+Pantheon+$H_0$
datasets. The reference model is the $\Lambda$CDM model. The left
panel is that for the TT spectrum, the right one is for the EE and the bottom one is for the TE spectrum.}
    \end{figure}

We also follow the Ref.\cite{Lin:2019qug} and plot the difference ${\Delta}C_l=C_{l,model}-C_{l,ref}$ in units of the cosmic variance per multipole
\[ \sigma_{CV} =\left \{
\begin{array}{rl}
	\sqrt{2/(2l+1)}C^{TT}_l,&TT\\
	\sqrt{1/(2l+1)}\sqrt{{C^{TT}_l}{C^{EE}_l}+(C^{TE}_l)^2},&TE\\
	\sqrt{2/(2l+1)}C^{EE}_l,&EE\\
\end{array}\right. \]
for both parametrizations in CMB TT, EE and TE spectrum in Fig.\ref{fig2},
where $C_{l,ref}$ is that for the ${\Lambda}$CDM model. Compared with the results in
Ref.\cite{Poulin:2018cxd}, the residual oscillations caused by EDE
at small scales are slightly amplified under both the CPL and
oscillating parametrizations. The residuals of TT, EE and TE spectrum are within the cosmological variance at obserable scales ($l\lesssim2000$).

\subsection{AdS-EDE model}
In AdS-EDE model \cite{Ye:2020btb}, we consider the potential as
$V(\phi)=V_0({\phi\over M_p})^{4}-V_{ads}$, which is glued to
$V(\phi)=0$ at ${\phi}=(\frac{V_{ads}}{V_0})^{1/4}M_p$ by
interpolation, where $V_{ads}$ is the depth of AdS well. The
existence of AdS phase enables the density $\rho_{ede}$ of field
dilutes away faster, and so allows a larger EDE fraction but
without spoiling fit to CMB and BAO data, which makes AdS-EDE
possible have a higher $H_0$.

\begin{figure}[htbp]
    {\includegraphics[width=0.5\textwidth]{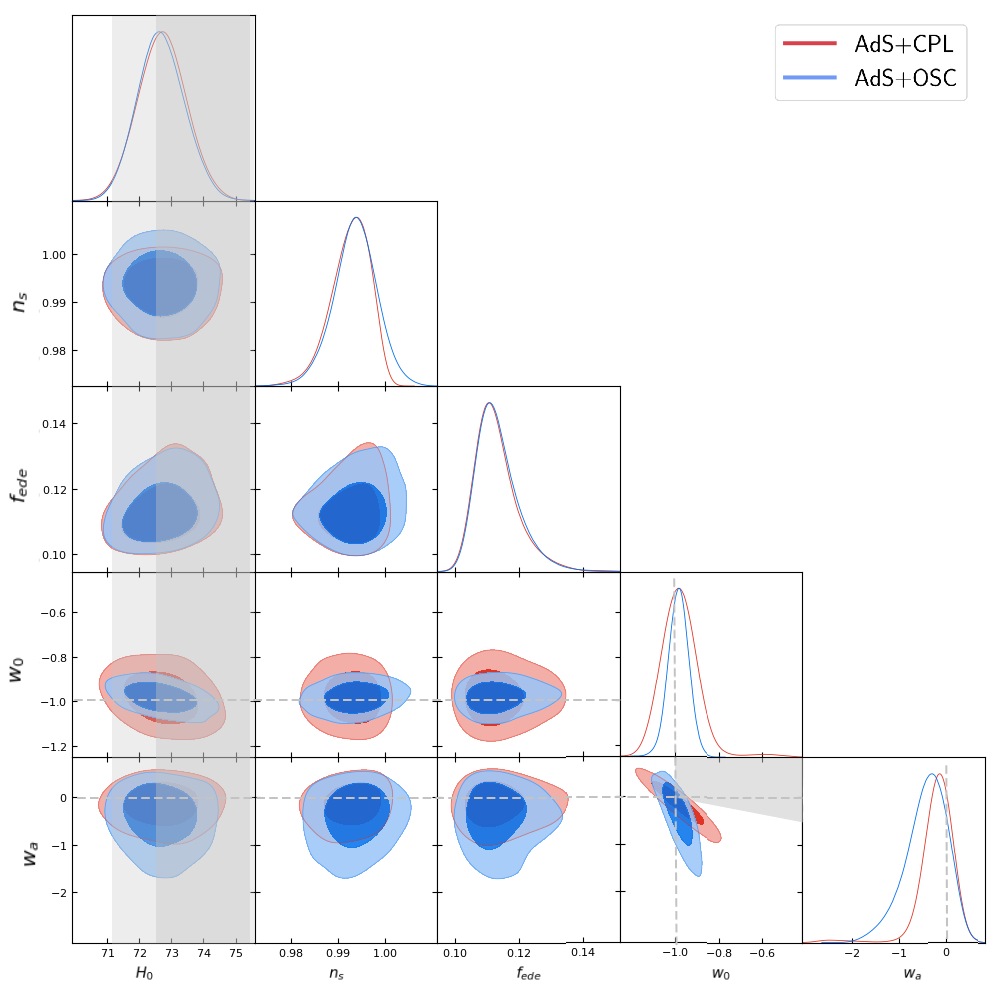}}\label{AdS}
\caption{\label{fig3}Marginalized $1\sigma$ and $2\sigma$ posterior
distributions of \{$H_0$, $n_s$, $f_{ede}$, $w_0$, $w_a$\} in
AdS-EDE model. The red lines are the results for the CPL
parameterization and the blue ones are for the oscillating
parameterization. The shadow regions are the same as in
Fig.\ref{fig2}.}
\end{figure}

We perform the MCMC analysis on the parameters set
$\{\omega_b,\omega_{cdm},H_0,\ln({10^{10}}A_s),n_s,\tau_{reio},\ln(1+z_c),
f_{ede},\alpha_{ads},w_0,w_a\}$, where $z_c$ is the redshift when
the field starts rolling, $f_{ede}$ is the energy fraction of EDE
at $z_c$, and $\alpha_{ads}$ corresponds to $V_{ads}$ by
$V_{ads}=\alpha_{ads}\lf(\rho_m(z_c)+\rho_r(z_c)\rt)$, which will
be fixed to $3.79\times10^{-4}$, see \cite{Ye:2020btb}. In
Table-II, we present the MCMC results of AdS-EDE model for the CPL
and oscillating parametrizations of $w(z)$, respectively. In
Fig.\ref{fig3}, we show the 1$\sigma$ and 2$\sigma$ marginalized
posterior distributions of model parameters.

\begin{table}[htbp]
    \begin{tabular}{c|c|cc}
        \hline
        Parameters&$\Lambda$CDM&AdS+CPL&AdS+OSC\\
        \hline
        100$\omega_b$&$2.247(2.224)^{+0.015}_{-0.014}$&$2.327(2.320)^{+0.0197}_{-0.0179}$&$2.326(2.324)^{+0.0184}_{-0.0192}$\\
        $\omega_{cdm}$&$0.1182(0.1183)^{+0.0008}_{-0.0013}$&$0.1353(0.1338)^{+0.00184}_{-0.00210}$&$0.1354(0.1354)^{0.00185}_{-0.00203}$\\
        $H_0$&$68.16(68.23)^{+0.56}_{-0.4}$&$72.77(72.30)^{+0.695}_{-0.805}$&$72.73(72.51)^{+0.778}_{-0.774}$\\
        ln($10^{10}$$A_s$)&$3.049(3.054)^{+0.013}_{-0.016}$&$3.071(3.067)^{+0.0158}_{-0.0151}$&$3.069(3.068)^{+0.0152}_{-0.0144}$\\
        $n_s$&$0.9688(0.9696)^{+0.0039}_{-0.0042}$&$0.9938(0.9951)^{+0.00488}_{-0.00468}$&$0.9934(0.9898)^{+0.00431}_{-0.00489}$\\
        $\tau_{reio}$&$0.0604(0.0636)^{+0.0066}_{-0.0075}$&$0.0530(0.0537)^{+0.00807}_{-0.00815}$&$0.0521(0.0523)^{+0.00772}_{-0.00753}$\\
        $f_{ede}$&-&$0.1137(0.1081)^{+0.00415}_{-0.00883}$&$0.1135(0.1094)^{+0.00772}_{-0.00753}$\\
        ln$(1+z_c)$&-&$8.168(8.169)^{+0.0647}_{-0.0848}$&$8.177(8.087)^{+0.0759}_{-0.0821}$\\
        \hline
        100$\theta_s$&$1.0422(1.0421)^{+0.0005}_{-0.0004}$&$1.0410(1.0410)^{+0.000304}_{-0.000308}$&$1.0410(1.0411)^{+0.000302}_{-0.000308}$\\
        $\sigma_8$&$0.8078(0.81)^{+0.0054}_{-0.0066}$&$0.8633(0.8636)^{+0.0112}_{-0.0106}$&$0.8651(0.8577)^{+0.0112}_{-0.0118}$\\
        \hline
        $w_0$&-1&$-0.987(-0.989)^{+0.0734}_{-0.0753}$&$-0.983(-0.960)^{+0.0463}_{-0.0477}$\\
        $w_a$&0&$-0.122(-0.154)^{+0.299}_{-0.248}$&$-0.427(-0.562)^{+0.567}_{-0.344}$\\
        \hline
    \end{tabular}
\caption{Mean values (best-fits) of parameters in AdS-EDE model
for CPL and oscillating parameterizations, respectively, fitted to
Planck2018+BAO+Pantheon+$H_0$ dataset. Actually, we can have the
bestfit $H_0\gtrsim 72$km/s/Mpc for AdS-EDE model without $H_0$
prior, see Appendix for the MCMC results based on
Planck2018+BAO+Pantheon dataset.}
    \label{tab1}
\end{table}

\begin{figure}[htbp]
   \subfigure{\includegraphics[width=0.45\textwidth]{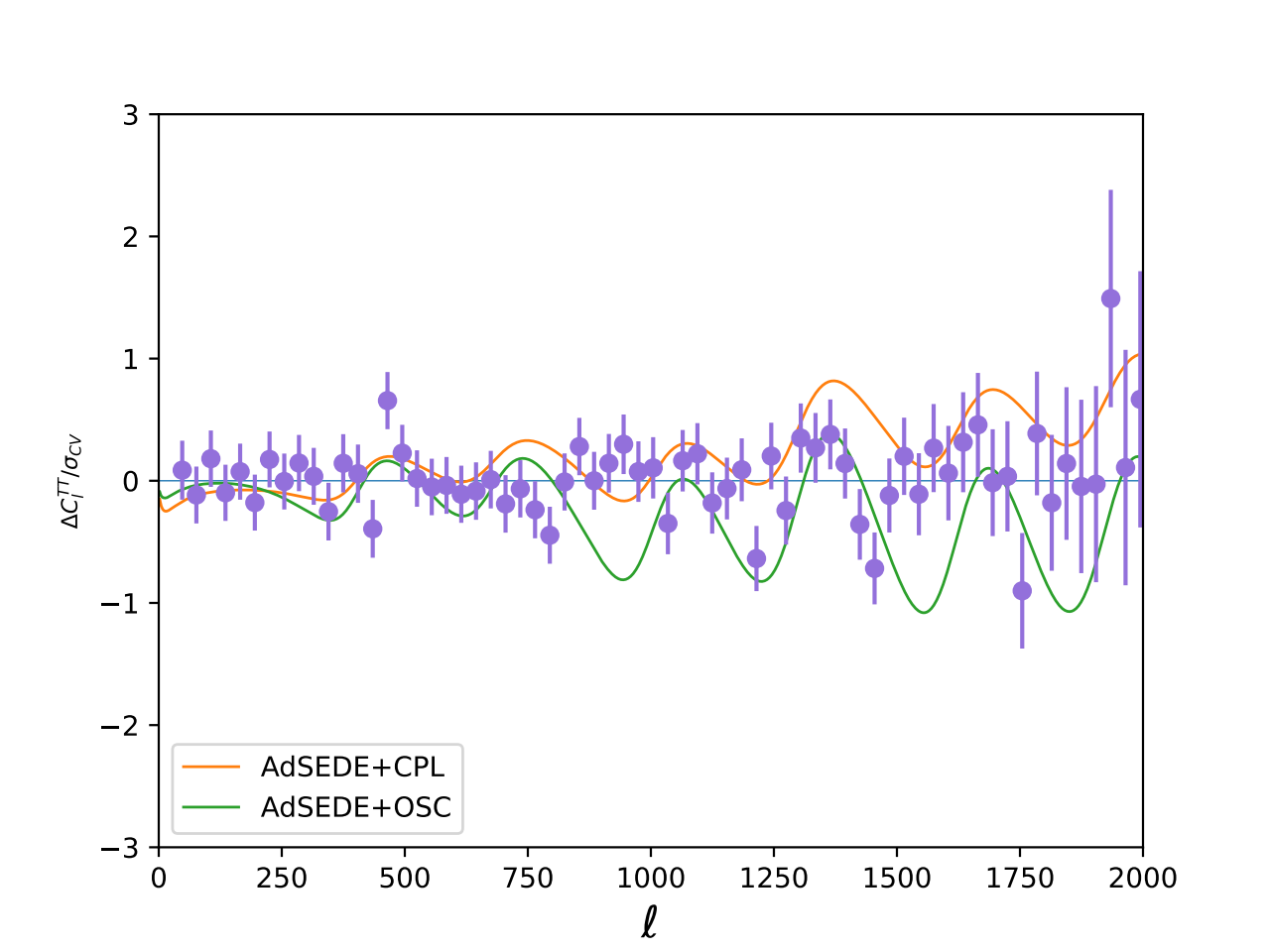}}\label{AdSTT}
    \subfigure{\includegraphics[width=0.45\textwidth]{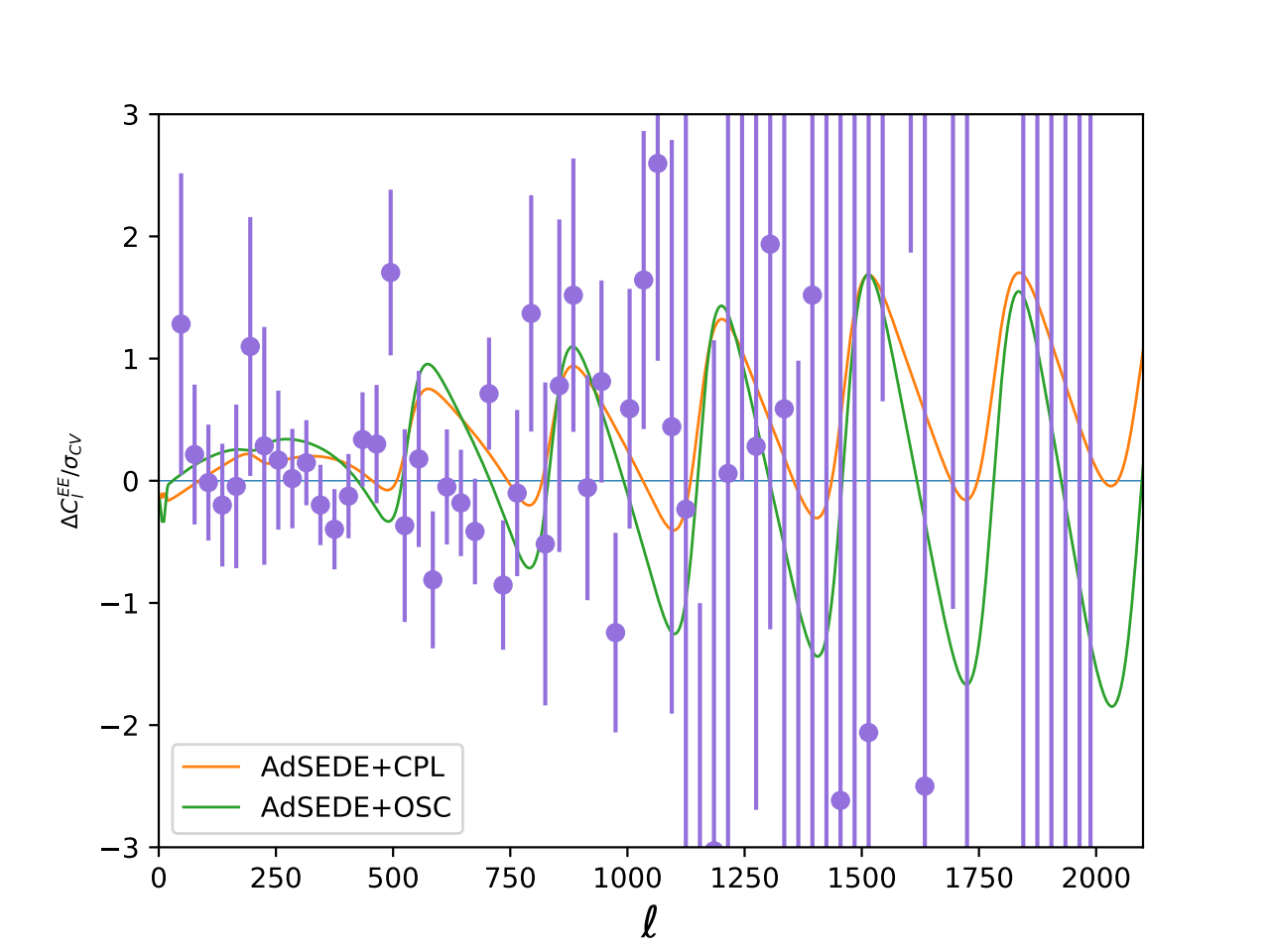}}\label{AdSEE}
    \subfigure{\includegraphics[width=0.45\textwidth]{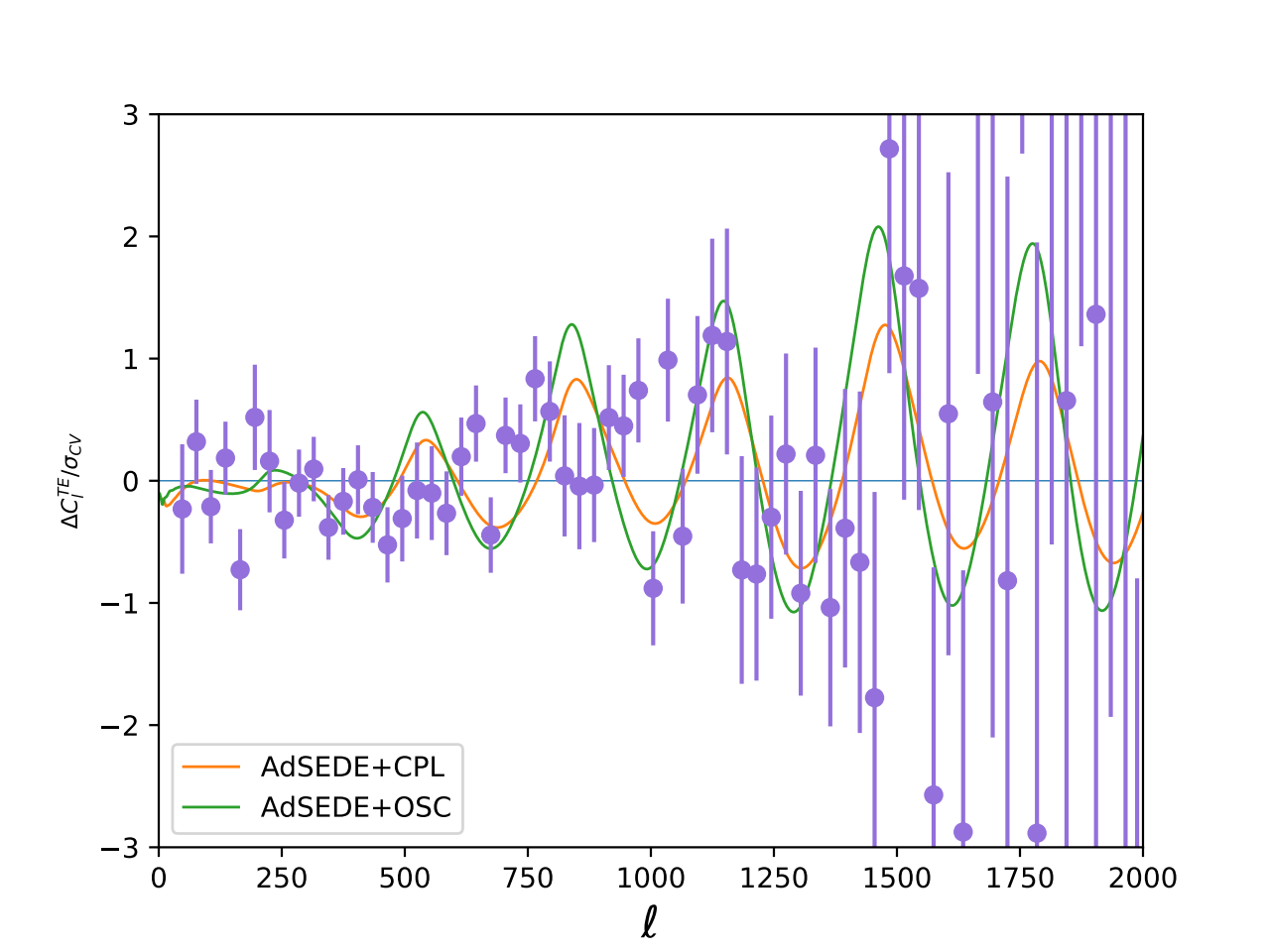}}\label{AdSTE}
\caption{\label{fig4}${\Delta}C_l/\sigma_{CV}$ for both parametrizations in
AdS-EDE model fitted to Planck2018+BAO+Pantheon+$H_0$ datasets.
The reference model is the $\Lambda$CDM model. The left panel is
that for the TT spectrum, the right one is for the EE and the bottom one is for the TE spectrum.}
\end{figure}

In Table-II and Fig.\ref{fig3}, we see again that the evolving $w$
has little effect on ${H_0}$ and $n_s$, compared with AdS-EDE
model with $w=-1$ in Ref.\cite{Ye:2020btb}. It is also clear that
the parameterizations of dark energy hardly affect the EDE
parameters. The result on $w(z)$ is still compatible with the
cosmological constant ($w_0=-1,w_a=0$), and only marginally favors
the evolution of $w$. The difference ${\Delta}C_l/\sigma_{CV}$ is
plotted in Fig.\ref{fig4}. Compared with the results in
Ref.\cite{Ye:2020btb}, the parameterizations of dark energy not
only maintain the character of oscillating in both TT and EE
spectrum, but also strengthen the going-upwards of amplitude with
$l$ in TT spectrum and the bump at $l\sim200$ in EE spectrum. The residuals of the TT spectrum are within the cosmological variance for large scales ($l\lesssim1500$) and become comparable to the $\sigma_{CV}$ as $l$ grows. However, the residuals of the EE and TE spectrum are larger than the $\sigma_{CV}$ in the $l\gtrsim1300$ multipoles for CPL and $l\gtrsim850$ for oscillating one, which may be detected.

\subsection{Discussion}

\begin{figure}[htbp]
        \subfigure[$CPL$]{\includegraphics[width=0.45\textwidth]{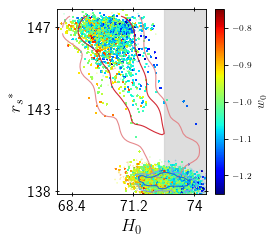}}\label{cplr}
       \subfigure[$OSC$]{\includegraphics[width=0.45\textwidth]{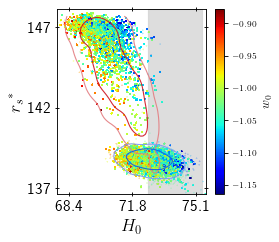}}\label{oscr}
\caption{\label{fig5}$H_0-r_s^*$ contours for the CPL and oscillating
parametrizations with colored scatters by $w_0$. The red contour
is that for Axion EDE model and the blue one is for AdS-EDE model.
The shadow region corresponds to 1 $\sigma$ value of local $H_0$
measurement.}
    \end{figure}

We present the $H_0-r_s^*$ contours for the CPL and oscillating
parametrizations, respectively, with colored scatters as $w_0$ in
Fig.\ref{fig5}. We see that $w_0$ at $1\sigma$ contour is closed
to -1, which is consistent with the cosmological constant. As
expected, we have $H_0\simeq 73$ and $w_0\simeq -1$ for AdS-EDE
model.

We list the $\chi^2$ of all datasets for different models in
Table-III, respectively. We find that all best-fit models are improved over the best-fit 
$\Lambda$CDM model by $\Delta\chi^2\sim{-20}$. We see that both parameterizations reduce
the $\chi^2$ of Axion EDE model markedly, but slightly reduce that
of the AdS-EDE model. This suggests that with the evolving $w$ of
current dark energy, Axion model seems to be favored over the AdS
model. However, here since the AdS parameter $\alpha_{ads}$
(relevant with the depth of AdS well) is fixed as
3.79$\times{10}^{-4}$, releasing $\alpha_{ads}$ might gives better
fit for the AdS-EDE model. It is also noted that although the
$\chi^2_{CMB}$ in Axion EDE model is reduced, its fit to BAO
dataset is worsened seriously.

    \begin{table}[htbp]
        \begin{tabular}{c|c|cc|cc}
            \hline
            Dataset&$\Lambda$CDM&AdS+CPL&AdS+OSC&Axi+CPL&Axi+OSC\\
            \hline
            Planck high-l TT,TE,EE&2347.5&2346.52&2344.81&2338.89&2336.30\\
            Planck low-l EE&398.2&396.04&395.77&395.73&396.29\\
            Planck low-l TT&23.9&20.63&20.81&20.94&20.78\\
            Planck lensing&9.1&10.07&10.81&10.33&9.36\\
            BAO BOSS DR12&1.8&3.51&3.51&4.61&5.08\\
            BAO smallz 2014&2.2&1.61&2.00&2.88&3.38\\
            Pantheon&1026.9&1028.24&1027.49&1027.12&1027.52\\
            HST&15.4&0.21&0.29&0.89&0.95\\
            \hline
            Total&3825&3806.83&3805.50&3803.40&3799.67\\
            \hline
        \end{tabular}
        \caption{$\chi^2$ of all datasets for different models.}
        \label{tab3}
    \end{table}

To test whether the smoothing effect of lensing to the CMB power
spectrum is consistent with that measured by the lensing
reconstruction, the lensing potential is often scaled by a
consistency parameter $A_L$, theoretically $A_L= 1$
\cite{Calabrese:2008rt}. It has been pointed out that Planck data
seems favor a closed universe \cite{DiValentino:2019qzk}, while
flat universe suggests $A_L=1.180\pm0.065$ (Planck TT,TE,EE+lowE)
\cite{Planck:2018vyg}, which is called the lensing anomaly. The
oscillating parameterization of dark energy might help to explain
this problem.

We set $A_L$ as a MCMC parameter, and show the posterior
distribution of parameters set $\{H_0,n_s,w_0,w_a,A_L\}$ in Axion-like
EDE model with the oscillating parametrization in Fig.6. We see
that the bestfit of $H_0$ is consistent with that in sect-III.A,
and $A_L=1.0421^{+0.036}_{-0.050}$, but with a slightly smaller
$\Omega_m$, see Fig.\ref{fig7}. Thus it is possible to seek for
certain oscillating parametrizations to alleviate the lensing
anomaly.

\begin{figure}[htbp]
    {\includegraphics[width=0.5\textwidth]{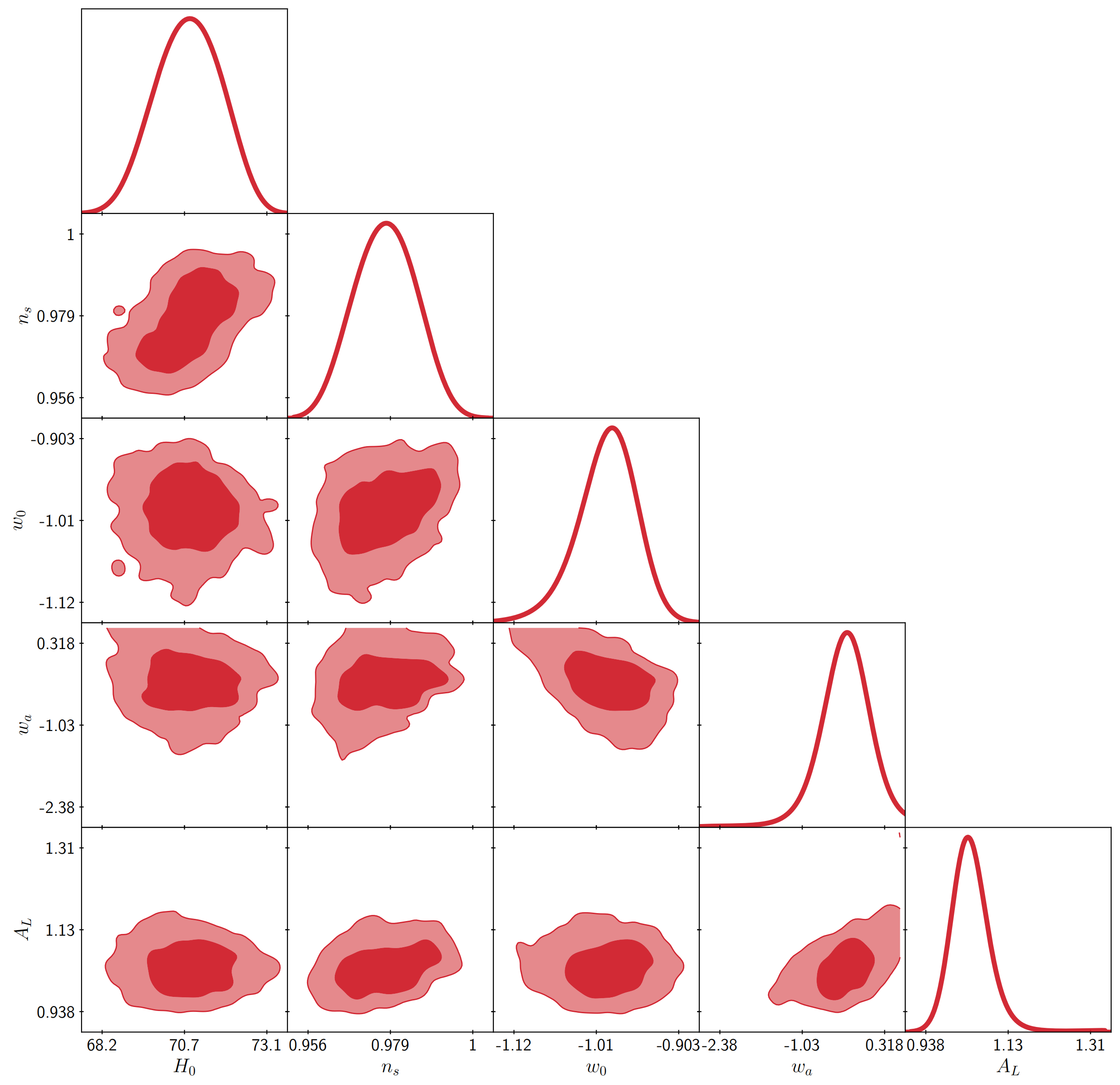}}\label{AL}
\caption{\label{fig6}Marginalized $1\sigma$ and $2\sigma$ posterior
distributions of \{$H_0$, $n_s$, $w_0$, $w_a$, $A_L$\} in Axion-like
EDE model with the oscillating parametrization. It is obvious that
$A_L$ is very closed to 1, which alleviates the lensing anomaly.}
\end{figure}
\begin{figure}[htbp]
    {\includegraphics[width=0.4\textwidth]{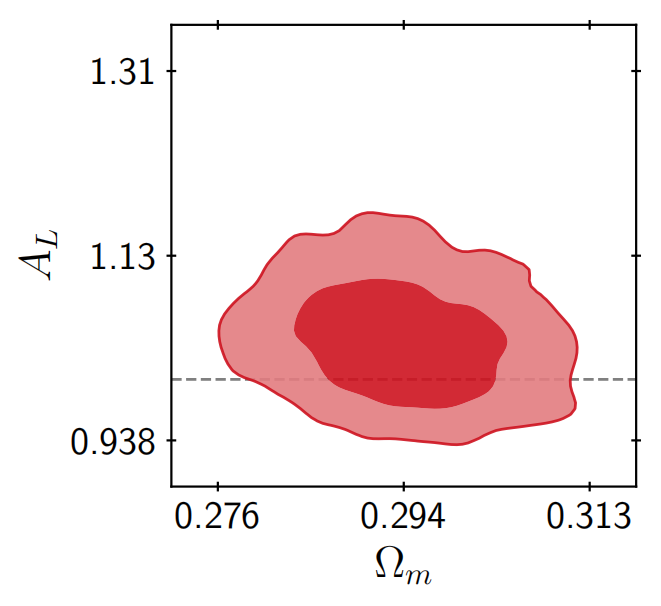}}\label{m}
\caption{\label{fig7}$A_L$-$\Omega_m$ contour in Axion-like EDE model with the
oscillating parametrization. The dash line corresponds to $A_L$=1.}
\end{figure}

\section{Conclusions}

In the studies on pre-recombination EDE, the evolution of Universe
after recombination is usually regarded as $\Lambda$CDM-like,
which corresponds to $w=-1$. However, in realistic models, $w$
might be evolving or oscillating. Here, we investigate the effects
of different parametrization of $w$ in Axion-like EDE and AdS-EDE
models, respectively.

We performed the MCMC analysis with recent cosmological data, and
found that bestfit $w(z)$ is compatible with  $w_0=-1,w_a=0$ (the
cosmological constant), and the evolution of $w$ is only
marginally favored. Particularly, our study confirmed again that
AdS-EDE model (here the bestfit $H_0\simeq 72.7$km/s/Mpc) is more
efficient in solving the Hubble tension. The parametrizations of
$w$ i.e.the evolution of current dark energy, in EDE models has
little effect on the bestfit values of ${H_0}$ and $n_s$, thus the
shift of primordial scalar spectral index scales still as ${\delta
n_s}\simeq 0.4{\delta H_0\over H_0}$ \cite{Ye:2021nej}, which
suggests a scale-invariant Harrison-Zeldovich spectrum ($n_s= 1$)
for $H_0\sim 73$km/s/Mpc, see also Table-I,II.

We also show ${\Delta}C_l/\sigma_{CV}$ for both parametrizations in
CMB TT, EE and TE spectrum, and found that compared with the results
in Refs.\cite{Poulin:2018cxd,Ye:2020btb}, the parameterization of
$w(z)$ basically maintains the shape of spectrum, but slightly
amplifies the residual oscillations caused by EDE at small scale,
which might be detectable. In addition, we also found that the
oscillating parametrization could alleviate the lensing anomaly.
Thus it is interesting to test other parametrizations of current
dark energy in EDE cosmologies.

\section*{Acknowledgments}

We thank Gen Ye, Jun-Qian Jiang for helpful discussions. This work
is supported by the NSFC, Nos.12075246, 11690021, the KRPCAS,
No.XDPB15, and also UCAS Undergraduate Innovative Practice
Project. We acknowledge the use of publicly available codes
AxiCLASS (\url{https://github.com/PoulinV/AxiCLASS}) and
classmultiscf
(\url{https://github.com/genye00/class_multiscf.git}).

\appendix
\section{MCMC results in AdS-EDE model without $H_0$ prior}
We show the MCMC results of AdS-EDE model for the CPL
parametrization, based on Planck2018+BAO+Pantheon dataset without
$H_0$ prior.
    \begin{table}[htbp]
        \begin{tabular}{c|c}
            \hline
            Parameters&AdS+CPL\\
            \hline
            100$\omega_b$&$2.331(2.328)^{+0.0194}_{-0.0210}$\\
            $\omega_{cdm}$&$0.1345(0.1340)^{+0.00186}_{-0.00227}$\\
            $H_0$&$72.44(72.36)^{+0.704}_{-0.925}$\\
            ln($10^{10}$$A_s$)&$3.070(3.075)^{+0.0156}_{-0.0167}$\\
            $n_s$&$0.9955(0.9935)^{+0.00462}_{-0.00488}$\\
            $\tau_{reio}$&$0.0537(0.0553)^{+0.00838}_{-0.00876}$\\
            $f_{ede}$&$0.1122(0.1059)^{+0.00446}_{-0.00909}$\\
            ln$(1+z_c)$&$8.1943(8.2317)^{+0.0844}_{-0.0943}$\\
            \hline
            100$\theta_s$&$1.0411(1.0410)^{+0.000328}_{-0.000349}$\\
            $\sigma_8$&$0.8584(0.8649)^{+0.0119}_{-0.0123}$\\
            \hline
            $w_0$&$-0.969(-0.979)^{+0.056}_{-0.040}$\\
            $w_a$&$-0.173(-0.146)^{+0.108}_{-0.174}$\\
            \hline
        \end{tabular}
\caption{Mean values (best-fits) of parameters in AdS-EDE model
for the CPL parameterizations, fitted to Planck2018+BAO+Pantheon
dataset. }
        \label{tab4}
\end{table}

\end{document}